\documentclass[twocolumn,showpacs,preprintnumbers,amsmath,amssymb,superscriptaddress,nofootinbib]{revtex4}

\usepackage{graphicx}
\usepackage{dcolumn}
\usepackage{bm}

\usepackage{pgfplots}
\usepackage{tikz}
\usetikzlibrary{calc,3d}

\usetikzlibrary{shapes.geometric}

\xdefinecolor{mygreen}{RGB}{0, 105, 0}
\xdefinecolor{mygrey}{RGB}{155, 155, 155}
\xdefinecolor{greenfibers}{RGB}{0, 65, 0}

\xdefinecolor{green32}{RGB}{128, 191, 66}
\xdefinecolor{pink8}{RGB}{177, 75, 94}

\newcommand{\vett}[1]{\mathbf{#1}}

\newcommand {\tr} {\mbox{\rm tr\,}}

\newcommand {\sgn} {\mbox{\rm sgn\,}}
{\left\lbrace\begin{array}{@{}l@{}}}%
{\end{array}\right.}

\begin{document}


\title{Morphing of Geometric Composites via Residual Swelling}

\author{Matteo Pezzulla}
\affiliation{%
Sapienza Universita` di Roma, via Eudossiana 18, I-00184 Roma, Italy
}%

\author{Steven A. Shillig}
\affiliation{%
Virginia Tech, Engineering Science and Mechanics, 222 Norris Hall, Blacksburg, VA, USA.
}%

\author{Paola Nardinocchi}
\affiliation{%
Sapienza Universita` di Roma, via Eudossiana 18, I-00184 Roma, Italy
}%

\author{Douglas P. Holmes}

\affiliation{
Department of Mechanical Engineering, Boston University, Boston, MA, 02215.
}%

\date{\today}

\begin{abstract}
Understanding and controlling the shape of thin, soft objects has been the focus of significant research efforts among physicists, biologists, and engineers in the last decade. These studies aim to utilize advanced materials in novel, adaptive ways such as fabricating smart actuators or mimicking living tissues. Here, we present the controlled growth--like morphing of 2D sheets into 3D shapes by preparing geometric composite structures that deform by residual swelling. The morphing of these geometric composites is dictated by both swelling and geometry, with diffusion controlling the swelling--induced actuation, and geometric confinement dictating the structure's deformed shape. Building on a simple mechanical analog, we present an analytical model that quantitatively describes how the Gaussian and mean curvatures of a thin disk are affected by the interplay among geometry, mechanics, and swelling. This model is in excellent agreement with our experiments and numerics. We show that the dynamics of residual swelling is dictated by a competition between two characteristic diffusive length scales governed by geometry. Our results provide the first 2D analog of Timoshenko's classical formula for the thermal bending of bimetallic beams -- our generalization explains how the Gaussian curvature of a 2D geometric composite is affected by geometry and elasticity.  The understanding conferred by these results suggests that the controlled shaping of geometric composites may provide a simple complement to traditional manufacturing techniques.
\end{abstract}

\maketitle

The continuous shape change during the growth and decay of biological structures is a constant presence within the natural world. These structures morph to accommodate an influx of new material, either growing due to an increase of mass to the system from an external source, or swelling from the absorption of excess fluid, such as water, caused by a change in humidity. Often, these morphological changes from swelling or growth result in geometries that enhance the biological structure's functionality~\cite{Boudaoud2010,Mirabet2011}, for instance the Venus flytrap's leaves snap closed after osmotically swelling, and this structural reconfiguration is essential for its nutrition~\cite{Forterre2005a}. Some of the most dramatic growth--induced deformations occur with slender structures, such as growing leaves~\cite{Marder2003}, wrinkling skin~\cite{Kucken2005}, and the writhing of tendril--bearing climbers~\cite{McMillen2002}.  Thin structures like these significantly deform to adopt nontrivial three dimensional shapes because they must bend to release their stretching energy~\cite{Klein2007}. The coupling between growth and large deformations presents an interesting opportunity for the morphing of synthetic structures, whereby if specific regions within a thin material can be prescribed to stretch the overall structure will adopt a new shape. Here, we demonstrate the controlled morphing of a thin structure by locally altering its intrinsic geometry; introducing a new class of \emph{geometric composite} structures. While composite materials combine constituents to enhance physical or chemical properties, geometric composites combine different intrinsic geometries to produce shapes that differ from the individual components. Shape change in these geometric composites is triggered by the growth--like swelling of one region due to residual fluid in the surrounding material. By tuning the underlying materials and geometry, we present a novel and straightforward means to directly morph elastic materials into various $3$D shapes. This {\em growth--actuated manufacturing} provides a natural complement to traditional reductive~\cite{Masuzawa2000} and additive manufacturing techniques~\cite{Kruth1998, Kruth2007}. 

The continuous shape change during the growth and decay of biological structures is a constant presence within the natural world. These structures morph to accommodate an influx of new material, either growing due to an increase of mass to the system from an external source, or swelling from the absorption of excess fluid, such as water, caused by a change in humidity. Often, these morphological changes from swelling or growth result in geometries that enhance the biological structure's functionality~\cite{Boudaoud2010,Mirabet2011}, for instance the Venus flytrap's leaves snap closed after osmotically swelling, and this structural reconfiguration is essential for its nutrition~\cite{Forterre2005a}. Some of the most dramatic growth--induced deformations occur with slender structures, such as growing leaves~\cite{Marder2003}, wrinkling skin~\cite{Kucken2005}, and the writhing of tendril--bearing climbers~\cite{McMillen2002}.  Thin structures like these significantly deform to adopt nontrivial three dimensional shapes because they must bend to release their stretching energy~\cite{Klein2007}. The coupling between growth and large deformations presents an interesting opportunity for the morphing of synthetic structures, whereby if specific regions within a thin material can be prescribed to stretch the overall structure will adopt a new shape. Here, we show the controlled morphing of a thin structure by locally altering its intrinsic geometry; preparing structures that are effectively geometric composites. While composite materials combine constituents to enhance physical or chemical properties, geometric composites combine different intrinsic geometries to produce shapes that differ from the individual components\cite{Klein2007,Kim2012a}. Shape change in these geometric composites is triggered by the growth--like swelling of one region due to residual fluid in the surrounding material. By tuning the underlying materials and geometry, we present a novel and straightforward means to directly morph elastic materials into various $3$D shapes. This {\em growth--actuated manufacturing} provides a natural complement to traditional reductive~\cite{Masuzawa2000} and additive manufacturing techniques~\cite{Kruth1998, Kruth2007}. 

Understanding the interplay among geometry, mechanics, and swelling is essential for the controlled morphing of thin structures. Geometrically, Gauss's \emph{Theorema Egregium} states that the Gaussian curvature of a sheet is dictated by the local distances between points, and this intrinsic geometry is conserved by isometries. Therefore, if a stimulus changes these local distances, thereby changing the sheet's Gaussian curvature, the disk will most likely bend into a three dimensional shape. The underlying geometry of a sheet may be programmed mechanically through confinement, or it can be altered chemically by locally swelling regions of the material.  For swelling to induce an out--of--plane deformation it must stretch the sheet in a non--homogenous manner, either via selective surface wetting~\cite{Holmes2011} or anisotropic material properties~\cite{Kim2012a}. To describe the shape of the resulting deformed structures, the F\"oppl--von K\'arm\'an equations were generalized to account for growth~\cite{Dervaux2009}, and employed to study the morphogenesis of a blooming lily~\cite{Liang2011}, leaves~\cite{Liang2009} and the buckling of swelling gels~\cite{Mora2006}. Building on these ideas of elasticity and growth, the theory of non--Euclidean plates~\cite{Klein2007, Efrati2009} was developed without any {\em a priori} assumptions regarding the structure's displacements. This approach provided a powerful way to design and study the morphing of thin structures such as the buckling of sheets~\cite{Gemmer2011,Kim2012b} and ribbons~\cite{Santangelo2009} and the three dimensional transformations of hydrogels into helices~\cite{Wu2013}.  Dias {\em et al.} presented an elegant study of the inverse buckling problem, where the growth patterns corresponding to some desired axisymmetric shapes were found explicitly~\cite{Dias2011}. In both models, an analytical prediction of a morphing structure's shape is often elusive, even for simple geometries and symmetric shapes. These concepts of differential growth have been investigated in the 1D case of morphoelastic rods, where instabilities such as the circumferential buckling of a growing cylinder occur~\cite{Moulton2011}. This work will build upon the contributions from differential geometry and elasticity to analytically predict the shape of 2D geometric composites that change shape in response to residual swelling.

\begin{figure}[t]
\includegraphics[scale=1]{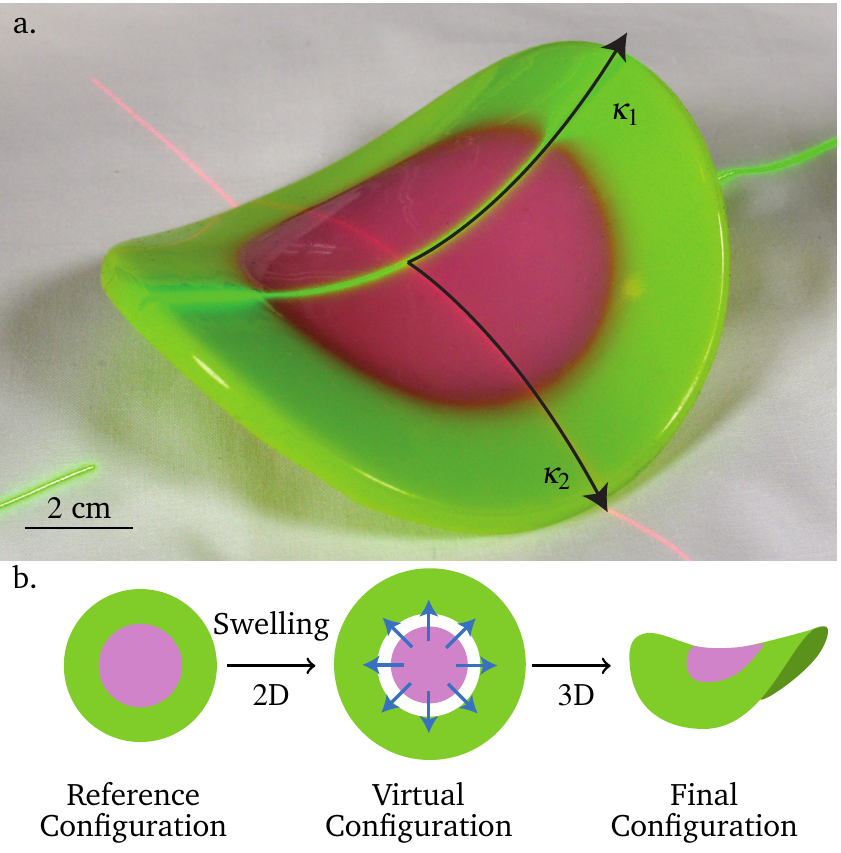}
\caption{(a) An initially flat, elastic plate morphs into a negatively curved shell, or saddle, in response to swelling of the annulus from residual free chains present in the inner disk. (b) A schematic representation of 2D plate morphing into a 3D shell.\label{figcartoon}}
\end{figure}

In Fig.~\ref{figcartoon} we show a geometric composite that morphs from a flat plate into a negatively curved surface from the residual swelling of the annulus (green) by free polymer chains present in the disk (pink). To predict this structure's shape change, we first constructed a mechanical analog based on the geometry of the geometric composite shown in figure~\ref{figcartoon}. In our mechanical analog, the sheet is made of a disk and an annulus that share the same material properties, but are geometrically incompatible: the inner radius of the annulus is slightly bigger or smaller than the radius of the disk so that the annulus has to be mechanically strained for the composite to be bonded together. The resulting geometrical frustration caused the disk to bend into positively or negatively curved surfaces, that is, into surfaces with positive or negative Gaussian curvature, respectively. This mechanical analog enforces the importance of geometry, and poses a common problem in elasticity usually cited as a prototypical example for frustrated structures, but has never been fully investigated~\cite{Efrati2009a}. In this work, we present an analytical model based on a membrane approximation that quantitatively captures the Gaussian curvature of the disks, and underscores the strong connection between the Gaussian curvature and the metric. The agreement among numerics, experiments, and analytics is excellent. Consequently, we apply our model to explain residual swelling, and study how the Gaussian curvature of the disk varies with the geometry of the composite. We show how the coupling between swelling and geometry dictates the main feature of structural morphing, and we provide the first $2$D extension of Timoshenko's formula for the thermal bending of bimetallic beams~\cite{Timoshenko1925}. Finally, we describe the dynamics of residual swelling, which exhibits a transition between two characteristic diffusive length scales depending on the geometry of the composite.

\section*{Intrinsic \& Extrinsic Geometry}
In thin structures, the effect of mechanical or chemical stimuli can be represented mathematically by a target metric~$\overline{a}$ -- a tensor that describes what the local distances between points should be in response to the applied stimuli\cite{Efrati2009}. The actual distances between points across a surface are represented by the realized metric $a$ and uniquely determine its Gaussian curvature, which describes the product of the surface's two principal curvatures, $\kappa_1$ and $\kappa_2$. If the Gaussian curvature imposed by the stimuli is not zero, the target metric cannot be realized in a flat disk, and the disk will bend to minimize the in--plane strains. Since the ratio of the bending to stretching energies scales like $\mathcal{U}_b/\mathcal{U_s}\sim h^2$, the disk will bend to minimize the in--plane strains if the thickness $h$ is small\footnote{The elastic energy is $\mathcal{U}=h\mathcal{U}_s+h^3\mathcal{U}_b$ where $\mathcal{U}_s$ has the dimension of an energy over a length, while $\mathcal{U}_b$ has the dimension of an energy over the cube of a length.}. Therefore, for thin disks, a reasonable approximation is to determine the structure's shape by minimizing the stretching energy, and allowing the disk to bend in whatever way necessary to accommodate the target metric. We write the bending energy density as $Eh^3(4H^2-K)$, where $H=(\kappa_1+\kappa_2)/2$ is the mean curvature of the mid--surface of the sheet, an extrinsic geometry property that depends on the embedding space, $K=\kappa_1\kappa_2$ is the Gaussian curvature, an intrinsic geometric property of the surface, and $E$ is the Young modulus of the sheet. Then, considering an incompressible material ($\nu=0.5$), the stretching energy may be written as~\cite{Efrati2009}
\begin{equation}
\label{eq-stretch}
\mathcal{U}_s\simeq h\int E[\tr^2(a-\overline{a})+\tr(a-\overline{a})^2] \sqrt{\lvert\overline{a}\rvert} \ \text{d}A\,.
\end{equation}
\section{Geometric Composites}
\subsection{Membrane Approximation}

Before we consider the dynamic problem of a swelling--induced changing metric, we construct as a proof of concept a mechanical analog of the geometric composite and consider the static case of a mechanically prescribed hyperbolic or elliptic metric over a disk. In the simplest case, when homothetic transformation radially stretches a circular disk, the radial and azimuthal distances all expand uniformly, and the disk stays flat since the surface's metric remains Euclidean (Fig.~\ref{fig:att}a). If we consider, instead, a circular disk of radius $R$ and an annulus of inner radius $R_i$ and outer radius $R_e$ (Fig.~\ref{fig:att}b), we can then impose a homothetic transformation independently on either the disk or the annulus. By defining $\alpha\equiv R_i/R$ as the mismatch between the disk and the annulus, it is immediately apparent that if $\alpha\neq1$ the two structures are incompatible and the annulus must be stretched ($\alpha<1$) or compressed ($\alpha>1$) to fit the circular disk (Fig.~\ref{fig:att}c). When the inner radius of the annulus is bonded to the edge of the circular disk, the resulting disk is a geometric composite that may be roughly modeled as a body having the following target metric in polar coordinates
\begin{equation}
\overline{a}=
f^2(r)
\begin{pmatrix}
1&0\\
0&r^2
\end{pmatrix}\,,
\quad    f^2(r) = \left\{
     \begin{array}{lr}
       1, & r\le R  \\
       \alpha^2, & r>R
     \end{array}
   \right.\,.
\end{equation}
This metric is flat within each of the two domains, but the metric of the geometric composite is not flat, {\em i.e.} there does not exist any parabola that fits $\overline{a}_{\theta\theta}(r)$, the azimuthal covariant metric coefficient. As in~\citenum{Audoly2003}, we approximate all the strains to zero but $a_{\theta\theta}-\overline{a}_{\theta\theta}$; therefore, if the disk and the annulus are made of the same material, the stretching energy from equation~\eqref{eq-stretch} reads (see {\it Appendix})
\begin{equation}\label{eq:func}
\mathcal{U}_s\simeq Eh\int_0^{R}\frac{(a_{\theta\theta}-r^2)^2}{r^{3}}\ \text{d}r+Eh\int_{R}^{R_e/\alpha}\frac{(a_{\theta\theta}-\alpha^2r^2)^2}{\alpha^{2}r^{3}}\ \text{d}r\,.
\end{equation}

Physical intuition tells that when the annulus is stretched (compressed), the disk will bend into a dome--like (saddle--like) shape. This statement may be mathematically represented as $\sgn K=\sgn(1-\alpha)$. To describe the resulting shape, we use Gaussian normal coordinates ($\rho$, $\theta$) to express the realized metric~\cite{doCarmo1976}, where $\rho(r)~=~\int_0^r\sqrt{a_{rr}(r')}dr'$ measures the arc length along radial geodesics while $\theta$ is the azimuthal angle. In these coordinates, the first fundamental form is written as $ds^2=d\rho^2+a_{\theta\theta}(\rho)d\theta^2$, and by the Gauss theorem, the Gaussian curvature is $-\partial_{\rho\rho}\sqrt{a_{\theta\theta}}/\sqrt{a_{\theta\theta}}$, where $\partial_{\rho\rho}$ is the second order partial derivative with respect to $\rho$. We minimize the stretching energy by looking for metrics with constant Gaussian curvature, that is $a_{\theta\theta}(\rho)=(\sin(\sqrt{K}\rho)/\sqrt{K})^2$.~\footnote{Notice that when the Gaussian curvature is negative, {\em i.e.} $K<0$, the metric may be rewritten as $a_{\theta\theta}(\rho)~=~(\sinh(\sqrt{-K}\rho)/\sqrt{-K})^2$.} As long as $\lvert K\rvert<1/R_e^2$~\footnote{This upper bound means that each principal direction cannot have a curvature that exceeds $1/R_e$.}, we can Taylor expand $a_{\theta\theta}(\rho)$ to linearize the metric in $K$ as
\begin{equation}\label{eq:att} 
a_{\theta\theta}(\rho)=\rho^2-\frac{K}{3}\rho^4+O(\rho^5)\,.
\end{equation}
Note that the first order term corresponds to a flat metric whereas the second one dictates the kind of non--Euclidean geometry that the disk will develop depending on the sign of $K$. The energy is quadratic in $K$ and therefore can be minimized analytically; notice that, if the annulus is neither deformed ($\alpha=1$) nor present ($R_i=R_e$), the energy is a simple parabola in $K$ with the minimum at $K=0$ since the disk is not constrained, and does not need to bend. Similarly, when the disk is not considered ($R=0$), the annulus does not need to bend either, and stays flat ($K=0$) with a radial stretch equal to $\alpha$. Once the stretching energy is minimized, we observe that the bending energy density is equal to $3H^2$ for dome--like shapes~($K=H^2$) and to~$4H^2-K$ for saddle--like shapes. In the latter case, since~$K<0$, the disk tries to morph into a minimal surface~($H=0$). Important aspects, albeit beyond the scope of this work, are the study of the discontinuities in the metric and the effect of a finite thickness.

When the target metric is elliptic, the resulting shape is unique \cite{Santangelo2009}. On the other hand, when the target metric is hyperbolic, the embedding is not unique, and shapes that are more complex then a saddle may develop when the thickness is very small\cite{Klein2011}. In our case, both experimental and numerical evidence indicate that the thickness to radius ratio ($\simeq 0.16$) is sufficiently high to avoid the development of complex shapes other than the saddle, yet small enough for the structure to be considered thin.

\section{Mechanical Analog}
\subsection{Experiments and Numerics}

To test our model, we prepared geometrically frustrated structures to realize dome--like and saddle--like disks, and measured their Gaussian curvatures. Circular molds were laser cut out of acrylic sheets, and used to cast samples with polyvinylsiloxane (PVS -- Zhermack Elite Double $32$). In these model experiments, we use an elastomer with a Young modulus $E=0.96$~MPa and a Poisson ratio $\nu=0.5$. The molds had a thickness $h=1.6$~mm with the radii of the disks varying between $5$ and $12$~mm. We designed the geometry so that the outer radius of the stretched disks was equal to $10$~mm. Bonding between the stretched annulus and the disk was accomplished by a small amount of uncrosslinked PVS. To measure their Gaussian curvature, we projected a laser sheet normal to the disk, and captured images of the reflected light with a Edmund--Optics GigE camera with a Nikkor lens ($35$~mm f: $1$-$1.4$) at $24$ equally spaced points along the disk's diameter. Image analysis was performed using Matlab to reconstruct the deformed shape. The annuli of polyvinylsiloxane elastomer were stretched homothetically and bonded to the inner disk. Upon release from the molds, the disks spontaneously morphed into domes or saddles. The annuli may be thought of as springs that want to release the energy by recovering their original shapes: for example, figure~\ref{fig:att}~(c) shows how the annulus must be compressed to get a saddle. We measured the shape of the deformed disks to determine the two principal radii of curvature, and plotted the Gaussian curvature in figure~\ref{fig:att}. These measurements confirmed the assumption used in our analysis that~$K$ is approximately constant across the disk. We also carried out numerical simulations to solve the problem within the context of finite incompressible tridimensional elasticity with large distortions using a Neo--Hookean material model~\cite{Lucantonio2014} implemented in the commercial software COMSOL Multiphysics. The stimulus in the numerical model is represented by a unimodular distortion field $\vett{F}_o=f(r)(\vett{I}-\vett{e_3}\otimes\vett{e_3})+f(r)^{-2}\vett{e_3}\otimes\vett{e_3}$, where $\vett{e_3}$ is the unit vector field orthogonal to the undeformed mid--surface of the disk. In this way, the tridimensional model is consistent with the $2$D model since $\vett{F}_o^T\vett{F}_o\cdot\vett{e_{\gamma}}\otimes~\vett{e_{\mu}}~=~\bar{a}_{\gamma\mu}$, where $\vett{e}_{\gamma}$ is the $\gamma$--th vector of the covariant basis spanning the undeformed mid--surface of the disk. The numerical results are also plotted in figure~\ref{fig:att}.

\subsection{Analysis}
\begin{figure}[!h]
\includegraphics[scale=1]{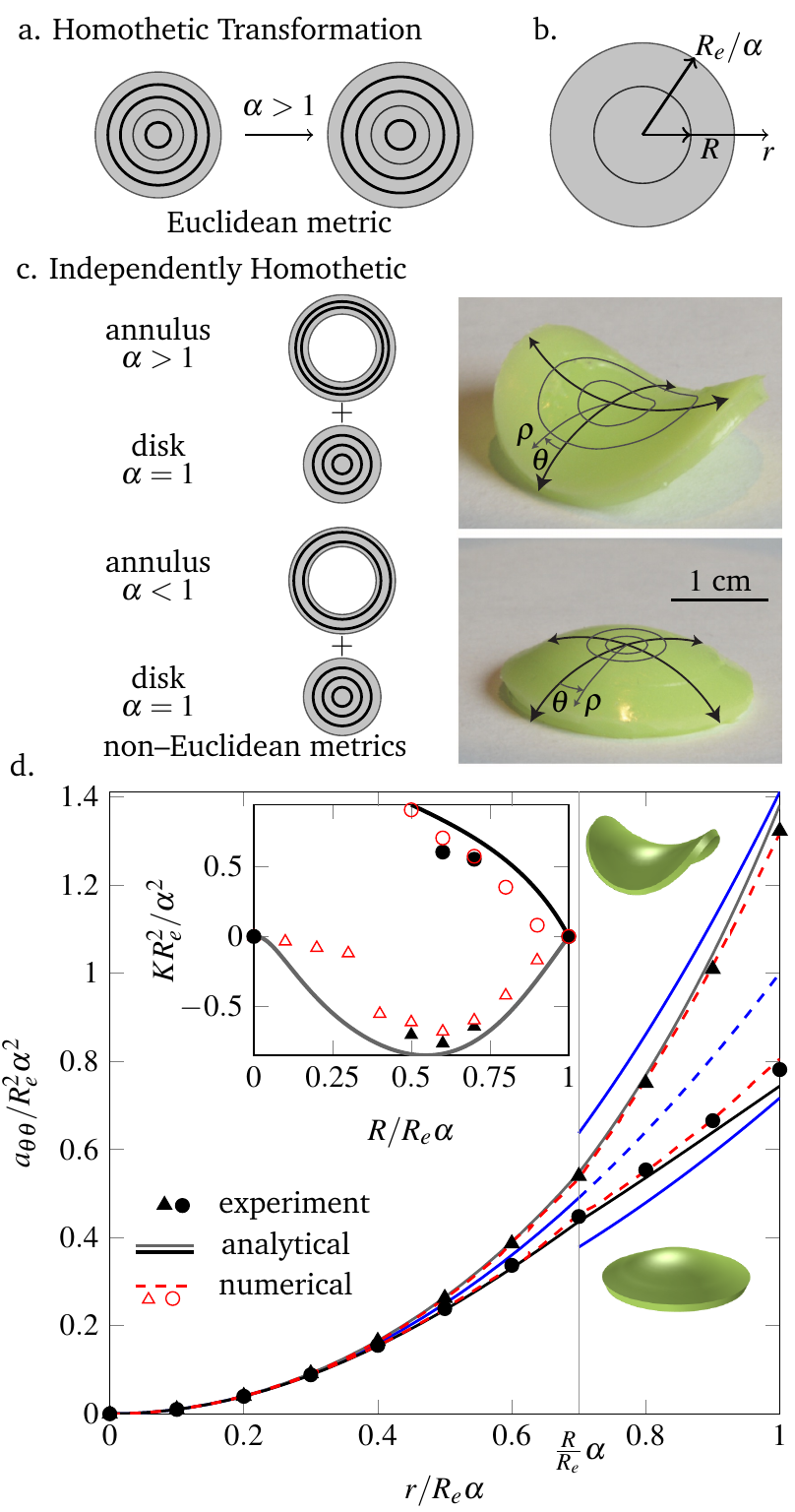}
\caption{(a) A schematic of the metric of a circular disk in polar coordinates changing by $\alpha$ as the disk is stretched homothetically. (b) A schematic of the relevant radii. (c) Schematics of the metrics of an annular ring and a circular disk stretched independently, along with images of the resulting shell denoting the principal directions of curvature and Gaussian normal coordinates. (d) A plot of the realized metric (analytical as solid black and grey curves, numerical as dashed red curves, experiments as black symbols) normalized by $R_e^2\alpha^2$ vs. the normalized radial coordinate for prescribed target metrics (blue curves) when $R/R_e\alpha=0.7$, which result in positively and negatively curved shells. The inset shows the normalized Gaussian curvature vs. the radii ratios (analytical as solid black and grey lines, numerical as red symbols, experiments as black symbols). \label{fig:att}}
\end{figure}

The main plot in figure~\ref{fig:att} refers to the case when $R/R_e\alpha=0.7$. The blue curves show the azimuthal component of the target metric of the mid--surface $a_{\theta\theta}$ as a function of $r$. As long as $r<R$, $a_{\theta\theta}$ is equal to $\rho^2$ since the inner disk has not been stretched; on the contrary, within the annulus ($r>R$), the azimuthal component is equal to $\alpha^2\rho^2$ with $\alpha<1$ ($\alpha>1$) if the annulus has been stretched (compressed). Notice that the parabola in the inner disk is represented also for $r>R$ (dashed blue curve) to show how the metric should look like if the disk were flat and not stretched. By the Taylor approximation of the metric, it is evident that if $K>0$ ($K<0$), $a_{\theta\theta}$ stays below (above) the parabola. Solid black lines are the analytical solutions of~\eqref{eq:func} for elliptic and hyperbolic target metrics. By looking at equation~\eqref{eq:func}, we notice that for the energy to be minimized, the analytical solution should be as close as possible (in a $L^2$ sense~\footnote{$L^2$ is the Lebesgue space of squared integrable functions.}) to the target metric with weight functions that are $r^{-3}$ and $\alpha^{-2}r^{-3}$ in the disk and in the annulus, respectively. This explains why the realized metric stays in between the target metrics, and closer to the target metric in the disk. Circles and triangles in the main plot show the experimental result (we measure $K$ and compute $a_{\theta\theta}$ from equation~\eqref{eq:att}). Dashed red curves represent the numerical solutions and show that the assumption of homogenous Gaussian curvature accurately describes the metric in this case apart from slight deviations near the edges. The experimental and numerical results are in excellent agreement with our closed form analytical solution. The numerics also shows that $H\simeq\sqrt{K}$ for domes and $H\simeq 0$ for saddles, as predicted analytically. It is interesting to note that the capability of estimating an extrinsic geometric quantity, \emph{i.e.} the mean curvature $H$, from a model built around intrinsic geometries allows us to determine the displacement field, {\em i.e.} the embedding of the structure, up to rigid motions. This indetermination does not affect the predictability of axisymmetric shapes like domes, but it does affect the one of saddles that have an axisymmetric metric but a not axisymmetric embedding. While we can predict the magnitude of the principal curvatures, the principal directions of curvature are dictated by imperfections in both experiments and numerics when the target metric is hyperbolic.

\subsection{Variation of Geometric Composition}
We then tested the models (analytical and numerical) for other values of $R/R_e\alpha$. When this ratio is between $0$ and $0.5$, the Gaussian curvature cannot be approximated as homogenous throughout the disk but attains two constant values in the inner disk and the annulus. However, the metric does not diverge much from the constant curvature solution, and the analytical model gives a result that is in good agreement with the mean Gaussian curvature of the disk, which is exactly what we measured experimentally. The inset in figure~\ref{fig:att} shows the agreement between experiments, numerics and analytics as~$R/R_e\alpha$ varies. Notice that the analytical model always overestimates the Gaussian curvature since it is based on a membrane approximation. The numerical model shows that, when the target metric is hyperbolic, wrinkles arise below a critical thickness as also shown in~\citenum{Klein2011}. We then expect our hypothesis of constant Gaussian curvature to hold in a finite range of thicknesses.

\section{Residual Swelling}
\subsection{Experiments}

Residual swelling is a fairly more complicated phenomenon than the geometrical confinement that dictated the shape change in the simplified mechanical problem. Swelling--induced deformations cannot be seen as distortions, as they are related to both the elastic properties of the gel, and the chemical conditions of the residual free polymer chains. Moreover, in this case swelling is driven by the concentration gradient of these chains across the entire structure. We used the circular molds of the mechanical analog to cast samples with polyvinylsiloxane as shown in figure~\ref{figcartoon} (PVS -- Zhermack Elite Double 32 for the annulus and Zhermack Elite Double 8 for the inner disk). Both elastomers are incompressible ($\nu=0.5$) and their Young's moduli were measured as $0.96$~MPa and $0.23$~MPa for PVS $32$ (annulus) and PVS $8$ (inner disk), respectively. The inner disks (radius $R$) and the annuli (radii $R$ and $R_e$) were geometrically compatible so that they could be bonded without pre--stretch. Once released from the molds, the geometric composites were flat, and the plates morphed into curved disks over time due to residual swelling -- the flow of free polymer chains from high density regions (softer gel, disk) to low density regions (stiffer gel, annulus). To study the influence of $R/R_e$ on the morphing process, we fixed the radius of the whole disk to $12$~mm and varied the radius of the inner disk from $5$~mm to $11$~mm casting $7$ disks with different~$R/R_e$. We measured the time evolution of the Gaussian curvature of each disk with the same procedure used for the mechanical analog, repeated every three hours.

\subsection{Residual Swelling of Geometric Composites}
\begin{figure}[!h]
\includegraphics[scale=1]{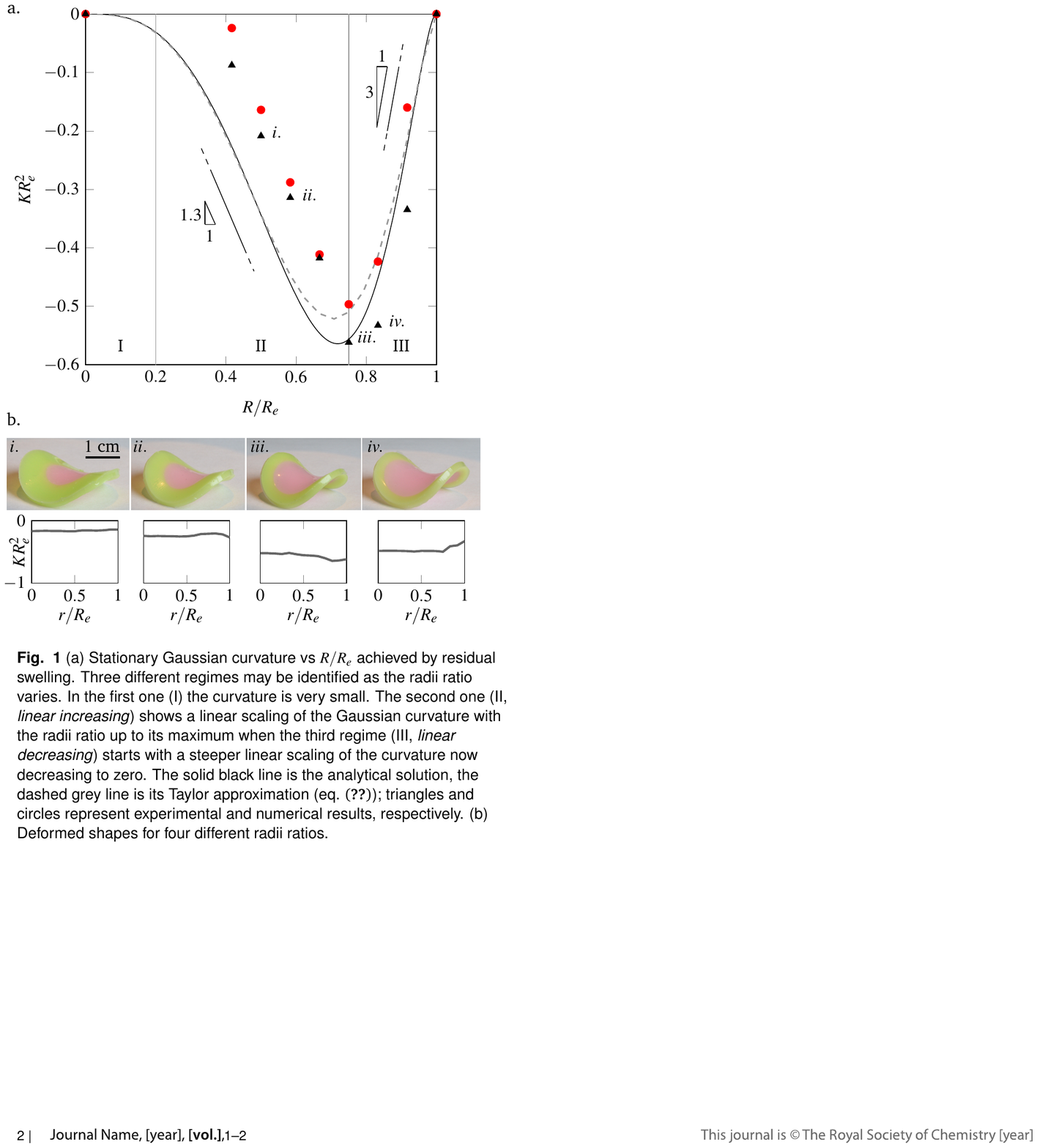}
\caption{(a) Stationary Gaussian curvature vs $R/R_e$ achieved by residual swelling. Three different regimes may be identified as the radii ratio varies. In the first one (I) the curvature is very small. The second one (II, \emph{linear increasing}) shows a linear scaling of the Gaussian curvature with the radii ratio up to its maximum when the third regime (III, \emph{linear decreasing}) starts with a steeper linear scaling of the curvature now decreasing to zero. The solid black line is the analytical solution, the dashed grey line is its Taylor approximation (eq.~\eqref{eq:Timo}); triangles and circles represent experimental and numerical results, respectively. (b) Deformed shapes for four different radii ratios and corresponding experimental profiles of the Gaussian curvature.\label{fig:KR}}
\end{figure}

While this problem couples nonlinear geometric mechanics with elastomer swelling, we can provide insight into this process by incorporating swelling into our mechanical analogy. The stretching ratio $\alpha$ now dictates the metric that each part of the disk would realize upon swelling if it were free (not bonded to the other). The inner disk and the annulus would like to shrink and swell, respectively, as molecules are flowing from the former to the latter. We assume that if the annulus would like to swell by a factor $\alpha$, the inner disk would like to shrink by a factor $\alpha^{-1}$.  Incorporating the difference between the two Young's moduli, the functional in equation~\eqref{eq:func} is modified as
\begin{equation}\label{eq:funcswelling}
\mathcal{U}_s\simeq\int_0^{R}\frac{(a_{\theta\theta}-\alpha^{-2}r^2)^2}{\alpha^{-2}r^{3}}\ \text{d}r+\frac{E_a}{E_d}\int_{R}^{R_e}\frac{(a_{\theta\theta}-\alpha^2r^2)^2}{\alpha^{2}r^{3}}\ \text{d}r\,.
\end{equation}
Notice that, since no pre--stretch is applied, the radius of the disk is $R_e$, \emph{i.e.} it coincides with the outer radius of the green annulus. The Young's moduli of the green annulus and the pink disk are denoted as $E_a$ and $E_d$, respectively.\footnote{Their ratio is roughly equal to $4$ and its variation with swelling is neglected.}  To analytically determine how $\alpha$ should vary with $R/R_e$, we denoted as $c_d$ and $c_a$ the concentrations of the diffusive species in the disk and in the annulus, respectively. Since molecules flowed from the disk to the annulus, we fixed $c_a<c_d$ and imposed the conservation of mass as $c_\textup{eq}\pi R_e^2=c_d\pi R^2+c_a\pi(R_e^2-R^2)$, where~$c_\textup{eq}$ denotes the concentration at equilibrium. Then, we reasoned that the stretching ratio $\alpha$ will be proportional to the cubic root of the mass uptake inside the annulus so that $\alpha^3-1\sim(c_\textup{eq}-c_a)\pi(R_e^2-R^2)$. Finally, by expressing $c_\textup{eq}$ from the mass conservation, we got
\begin{equation}\label{eq:alpha}
\alpha=\Biggl[1+\eta\left(c_d-c_a\right)\left(\frac{R}{R_e}\right)^2\left(1-\left(\frac{R}{R_e}\right)^2\right)\Biggr]^{1/3}\,,
\end{equation}
where $\eta$ is a proportionality coefficient having the dimension of the inverse of a concentration and representing the link between mass uptake and stretch.\footnote{Similar to the hydrophilicity coefficient introduced in~\citenum{Nardinocchi2013b} to describe stretching induced by cation's motion in ionic polymer--metal composites.} The presence of a concentration gradient of polymer chains with a polydisperse molecular weight makes identifying this parameter difficult, and beyond the scope of this work. Qualitatively, the bigger the free chains, the higher $\eta$ should be. Notice that $\alpha$ is equal to $1$ when the mass uptake is zero, that is when the structure is homogeneous ($R/R_e=0\ \text{or}\ 1$). This is the important difference with the mechanical analog, since~$\eta$ is not known for swelling as it depends on the material and chemical properties of the elastomers; we therefore used it as a fitting parameter. Figure~\ref{fig:KR} shows the stationary values of the Gaussian curvature of the seven disks after residual swelling as obtained in the experiments (triangles), numerics (circles) and analytics (solid curve). Numerical and analytical results are obtained by using~\eqref{eq:alpha} in~\eqref{eq:funcswelling} and setting $\eta(c_d-c_a)$ equal to~$0.54$ that sets $\alpha_\textup{max}=1.043$ from~\eqref{eq:alpha}. The three linear regimes identified in figure~\ref{fig:KR} point out that the maximum of the Gaussian curvature is not attained for $R/R_e=1/2$ but for $R/R_e\simeq0.77$. A similar result was obtained for the mechanical analog as can be seen in the inset in figure~\ref{fig:att}. These two observations let us conclude that both the dimensionality of the swelling and simple geometry shift the maximum of the Gaussian curvature to high radii ratios instead of $R/R_e=1/2$. By the conservation of mass, it can be demonstrated that if the swelling had been $1$D, the maximum mass uptake would have been attained at $R/R_e=1/2$; if it had been $3$D, the maximum would have been attained at $R/R_e=1/2^{1/3}\simeq0.8$. In our $2$D case, the maximum is attained when $R/R_e=1/\sqrt2\simeq0.7$ as also experiments showed. So, in general, if $n$ is the dimensionality of the swelling, the maximum mass uptake is achieved for $R/R_e=1/2^{1/n}$. The agreement among experiments, numerics and analytics is quite good, and it is remarkable that the analytical model captures the linear regimes with the same slopes. The closed form analytical solution of the problem is cumbersome, but it may be simplified by noticing that $\alpha_\textup{max}\simeq1$, which allows us to perform a Taylor expansion in terms of $\alpha_\textup{max}-1$. At the leading order, defining $\bar{E}=E_a/E_d$ and $\bar{R}=R/R_e$, it reads
\begin{equation}\label{eq:Timo}
KR_e^2\simeq96(1-\alpha_\textup{max})\bar{E}\bar{R}^3\frac{\bigl(1-\bar{R}^2\bigr)\bigl(1-\bar{R}^3\bigr)}{\bar{R}^6\bigl(1-\bar{E}\bigr)+\bar{E}}\,,
\end{equation}
which we think can be interpreted as the $2$D analog of Timoshenko's formula for beams~\cite{Timoshenko1925} as it expresses how the dimensionless Gaussian curvature varies with material and geometric ratios (see {\it Appendix}). To the best of our knowledge, this is the first analytical formula relating the Gaussian curvature of a geometric composite to its material and geometrical properties, \textit{i.e.} moduli and radii ratios. This simplified expression is represented in figure~\ref{fig:KR} as a grey dashed line, and is very close to the full solution: the linear regimes highlighted in the figure are in excellent agreement with our Timoshenko--like formula. The physical interpretation of our first order Taylor approximation is that the strains are assumed to be small, which is the same limit that Timoshenko obtained his formula within. It is worth noting that, unlike thermal stretches in uniformly heated bimetallic strips, the stretching ratio $\alpha$ should depend on the elastic properties of the geometric composites, as discussed in~\citenum{Lucantonio2014a}. 

\subsection{Swelling Dynamics}

\begin{figure}[!h]
\includegraphics[scale=1]{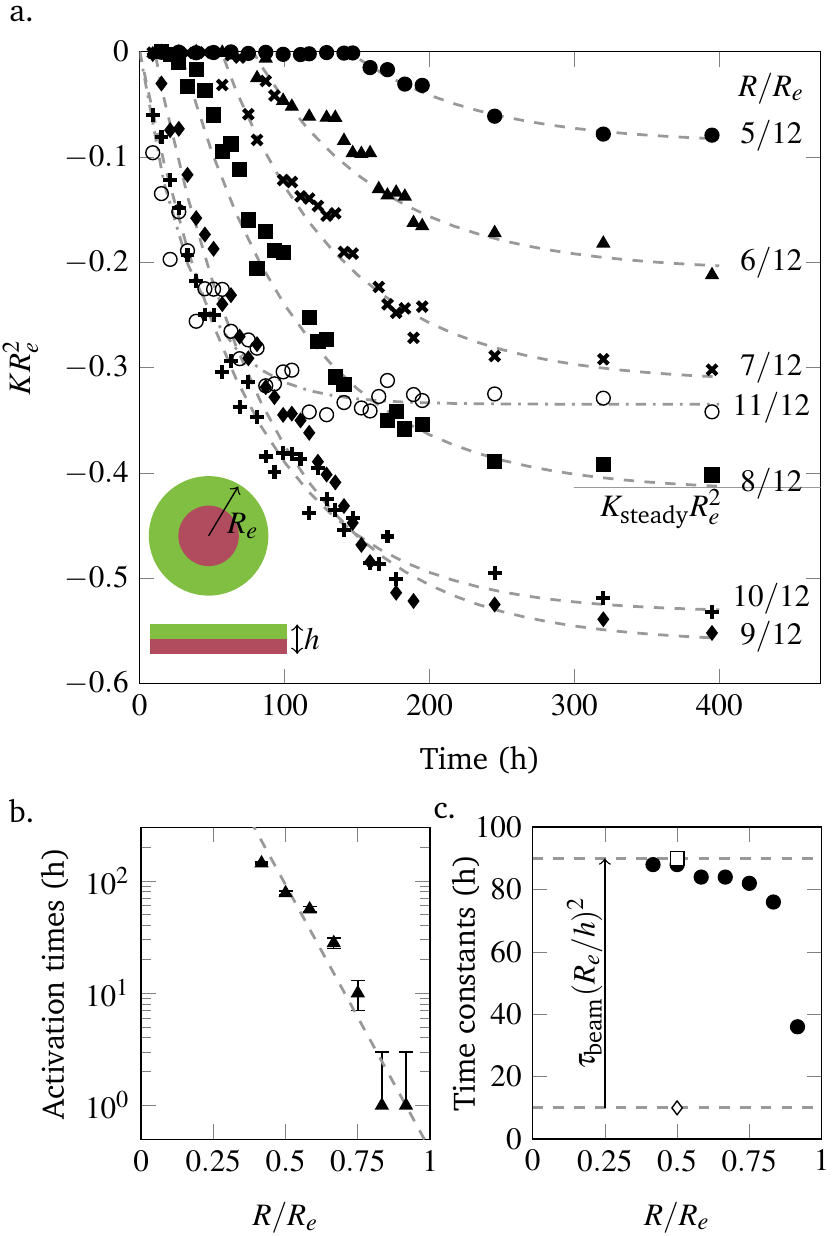}
\caption{(a) Normalized Gaussian curvature versus time for the different radii ratios. (b) Activation times versus $R/R_e$ (triangles) and numerical fitting (dashed curve). (c) Time constants of the disks versus $R/R_e$ (circles), time constant of the beam (diamond) and analytical estimate of the time constant of the disk corresponding to $R/R_e=1/2$ (square) assuming $\ell\sim R_e$. \label{fig:Kt}}
\end{figure}

 Our model successfully captures the steady--state morphology of residually swollen plates. Unlike the mechanical analog presented earlier, the residual swelling process adds a time--dependency to the deformation. The experimental results in figure~\ref{fig:Kt}~(a) show the time evolution of the Gaussian curvature of disks with seven different $R/R_e$, and the shape evolution contains two notable features: 1.) there is a critical \emph{activation time}, {\em i.e.} the time it takes for the structure to start deforming that depends on $R/R_e$, and 2.) following actuation, the disks deform in a diffusive manner. 

We assume that the swelling dynamics may be described as a diffusive process with a Fourier--like differential equation. The main features of a diffusive equation are that it is of the first order in time, giving rise to transients that are described by the exponential of time up to the steady--state, and of the second order in space (quasi--1D in our case). As we are studying the transient by looking at a homogeneous field -- the Gaussian curvature $K$ -- we focus on its variation with time, and note that the dashed lines in figure~\ref{fig:Kt}~(a) correspond to an exponential of time ($K_\textup{steady}R_e^2(1-e^{-t/\tau})$), as expected. Figure~\ref{fig:Kt}~(b) shows that the activation time varies with $R/R_e$ as $t_a\sim Ae^{-BR/R_e}$, where $A$ and $B$ are positive real numbers equal to $21038$~h and $10.862$, respectively, in our case.\footnote{Error bars correspond to $\pm3$~h since we measured $K$ every three hours.}  This numerical fitting is shown in the plot as a straight dashed grey line. Following activation, the diffusive shape change is characterized by the time constant~$\tau$ that dictates the time scale of the transient as it is the time at which the Gaussian curvature reaches the~$63\%$ of its stationary value. Figure~\ref{fig:Kt}~(a) shows that the disk corresponding to $R/R_e=11/12$ is faster than the other geometries; indeed, we found that its time constant is approximately $\tau\simeq40$~h whereas all the other disks have $\tau\simeq90$~h as shown in figure~\ref{fig:Kt}~(c).


As this is a diffusive process, we expect the dynamics will scale with the square of the characteristic length scale in the problem. We believe the observed difference in dynamics is the result of a change in the relevant characteristic length scale, \textit{i.e.} from the total radius of the disk to the width of the annulus. The characteristic time scale (time constant) of a diffusive process is equal to $\tau=\ell^2/D$, where $\ell$ is the characteristic length and $D$ is the diffusivity. While the latter is a property of the materials, the former is dictated by geometry. To identify the characteristic length, we examined a simpler geometry -- a bilayer beam. 
Using the same materials as in section 3.1,  we prepared a bilayered beam of equivalent thickness (figure~\ref{fig:Kt}~a - inset), and measured the time it took equilibrate into an arch, finding $\tau_{beam}\simeq 10$~h represented by a diamond in figure~\ref{fig:Kt}~(c). In this case, a reasonable assumption for the characteristic length is $\ell\sim h$, where $h$ is the total thickness of the beam as shown in the inset of figure~\ref{fig:Kt}~(a). Since the materials for the beam and disk are the same, they share the same value of~$D$. Therefore, we compared the $1$D diffusion in the beam with the $2$D diffusion in the disk with $R/R_e=1/2$ and assumed $\ell\sim R_e$ obtaining $\tau_\textup{disk}=(R_e/h)^2\tau_\textup{beam}\simeq90$~h: this analytical estimate is shown in figure~\ref{fig:Kt}~(c) as a square and it excellently predicts the experimental time constant of the disk. The experimental data show a decay of the time constant as the radii ratio approaches~$1$ that we interpret as the result of a decaying characteristic length, which represents the portion of the radius where swelling is actually taking place. When $R/R_e\simeq0.5$, our approximation $\ell\sim R_e$ is a good estimate for the characteristic length but when $R/R_e\rightarrow1$ the inner area of the disk is shielded from swelling and the characteristic length is smaller than $R_e$. We suggest that, as $R/R_e\rightarrow1$, the characteristic length approaches the width of the annulus.

\section{Conclusions} 
We have studied the morphing of geometric composites from flat plates into curved, three--dimensional shapes. The geometric composites morph by residual swelling, a phenomenon that we gained insight into by considering a mechanical analog that copies their geometry and morphs by geometrical confinement. The morphing problem of the mechanical analog is purely geometrical, and we developed an analytical model following the theory of non--Euclidean plates, which quantitatively describes how the Gaussian curvature is dictated by geometry. The strength of the model is indeed its analytical tractability that results from the assumption of a homogeneous Gaussian curvature throughout the disk. The agreement among experiments, numerics, and analytics is excellent even when the Gaussian curvature is not homogeneous because the analytical model provides a mean Gaussian curvature as a result, which is important for the design of actuators. 

We then employed the analytical model of the mechanical analog to study the morphing of geometric composites by approximating the swelling as a distortion. By using the conservation of mass, we analytically determined how the mass uptake should vary with the radii ratio, and assumed a linear proportionality between the mass uptake and the volume variation. The agreement among experiments, numerics and analytics is quite good and each approach identified three regimes for the Gaussian curvature as a function of the radii ratio: it is remarkable that the model catches these regimes and their linear features. Then, by assuming small stretches ($\alpha\simeq1$), we simplified the cumbersome analytical solution and provided the first $2$D extension of the Timoshenko's formula for beams. Finally, we studied the swelling dynamics and identified two different characteristic lengths depending on geometry. 

We think that the proposed model improved the understanding of the complex interplay among geometry, mechanics, and swelling. Additionally, the experiments demonstrate a robust and scalable means for the growth--actuated manufacturing of elastic shells -- a material that is traditionally difficult to prepare via additive and reductive manufacturing techniques. It is important to note that while residual fluid within the crosslinked elastomer drives the diffusion and swelling of the structure, the material behaves like an elastic solid, rather than a swollen gel. We expect this experimental procedure to translate to any combination of material--compatible elastomers where a gradient of small molecules can be programmatically prescribed. This may provide the foundation for an inkjet--like approach to 3D printing whereby small molecule fluids can be locally applied to a flat elastic sheet, allowing controlled diffusion to dictate the resulting growth pattern. Careful selection of the initial geometry will allow this technique to be used for generating regions of high curvature -- or folds -- which may form the basic building blocks for the growth of origami structures.

\section*{Acknowledgments}
We are grateful to Anupam Pandey (University of Twente) and Alexander Kotsch (Virginia Tech) for characterizing the experimental behavior of the bilayered beams, which provided helpful insights to the problem discussed herein. S.A.S. and D.P.H. acknowledge the financial support from NSF through CMMI-1300860. M.P. acknowledges the National Group of Mathematical Physics (GNFM-INdAM) for support (Young Researcher Project).

\appendix

\section{Derivation of the stretching energy of the mechanical analog}

Equation~\eqref{eq:func} is derived from~\eqref{eq-stretch} by computing the traces of the tensors $a-\bar{a}$ and $(a-\bar{a})^2$ using their polar coordinates. By definition, the trace of a tensor is the result of the contraction between the tensor and the metric tensor. As an example, the trace of $a-\bar{a}$ may be computed in polar coordinates as
\[
\tr(a-\bar{a})=\bar{a}^{\alpha\beta}(a_{\alpha\beta}-\bar{a}_{\alpha\beta})\,,
\]
where $\bar{a}^{\alpha\beta}$ denotes the contravariant components of the target metric tensor that may be thought as the entries of the inverse of the matrix $[\bar{a}_{\alpha\beta}]$. So, $\bar{a}^{rr}=f^{-2}(r)$, $\bar{a}^{r\theta}=0$ and $\bar{a}^{\theta\theta}=f^{-2}(r)r^{-2}$. Simple algebra yields:
\[
\tr(a-\bar{a})=\bar{a}^{\alpha\beta}a_{\alpha\beta}-2=f^{-2}(r)a_{rr}+f^{-2}(r)r^{-2}a_{\theta\theta}-2\,.
\]
Then, if $a_{rr}=\bar{a}_{rr}$, we get
\[
\tr^2(a-\bar{a})=(f^{-2}(r)r^{-2}a_{\theta\theta}-1)^2=\left(\frac{a_{\theta\theta}}{\bar{a}_{\theta\theta}}-1\right)^2\,.
\]
In our case, the trace of $(a-\bar{a})^2$ turns out to be equal to $\tr^2(a-\bar{a})$ so that
\[
\tr^2(a-\bar{a})+\tr(a-\bar{a})^2=2\left(\frac{a_{\theta\theta}}{\bar{a}_{\theta\theta}}-1\right)^2\,.
\]
Finally, to compute the stretching energy, we have to evaluate~$\sqrt{|\bar{a}|}$, that is
\[
\sqrt{|\bar{a}|}=f^2(r)r\,.
\]
By using the last two formulas, equation~\eqref{eq:func} is recovered after some rearrangements, discarding a factor $2$ since we are just interested in the minimization of the energy. 

\subsection{Gaussian normal coordinates} 
As $a_{rr}\simeq\bar{a}_{rr}$, the Gaussian normal coordinate $\rho$ is computed as
\[
\rho(r)\simeq\int_0^r\sqrt{\bar{a}_{rr}(r')}dr'=
\left\{
     \begin{array}{lr}
       r, & r\le R  \\
       R+\alpha(r-R), & r>R
     \end{array}
   \right.\,.
\]

\subsection{Gauss's Theorem} 
Since the polar coordinates are orthogonal, the Gauss's Theorem may be written as
\[
K=-\frac{1}{\sqrt{a_{\rho\rho}a_{\theta\theta}}}\Biggl[\Biggl(\frac{(\sqrt{a_{\theta\theta}}),_{\rho}}{\sqrt{a_{\rho\rho}}}\Biggr)_{,\rho}+\Biggl(\frac{(\sqrt{a_{\rho\rho}}),_{\theta}}{\sqrt{a_{\theta\theta}}}\Biggr)_{,\theta}\Biggr]\,.
\]
Notice that, since we are using Gaussian normal coordinates, $a_{\rho\rho}=a_{rr}(\partial r/\partial\rho)^2\simeq\bar{a}_{rr}f^{-2}(r)=1$ and $a_{\theta\theta}$ is the only metric coefficient contributing to the Gaussian curvature.

\section{Timoshenko--like formula}
Timoshenko wrote his formula for the bending of bi--metal thermostats as\cite{Timoshenko1925}
\[
\kappa h=\frac{6\Delta\varepsilon(1+m)^2}{3(1+m)^2+(1+mn)(m^2+\frac{1}{mn})}\,,
\]
where $\kappa$ is the curvature of the beam, $h$ is its thickness, $\Delta\varepsilon$ is the difference between the thermal strain in the two layers, $m$ is the ratio between the thicknesses of the two layers and $n$ is the Young moduli's ratio. Equation~\eqref{eq:Timo} resembles Timoshenko's formula in that it expresses a dimensionless curvature, albeit a Gaussian curvature, as a function of geometric and material ratios. Notice that the characteristic length is the one along which the composite is non homogeneous: the total thickness in the beam and the total radius in the disk, which may be thought of being obtained by rolling the bilayer beam.

\end{document}